\documentclass[preprint,prd,aps,showpacs,showkeys,nofootinbib]{revtex4}
\usepackage{amsmath}
\usepackage{graphicx}
\usepackage{dcolumn}
\usepackage{bm}
\usepackage{float}
\usepackage{color}
\usepackage{CJK,upgreek,fancyhdr}
\usepackage{bm}
\usepackage{diagbox}
\UseRawInputEncoding
\begin{document}

%\begin{CJK*}{GBK}{song}

%\preprint{hep-ph/0412147}

\title {A Regularized  $(XP)^2$ Model}
\author{Yu-Qi Chen\footnote{ychen@itp.ac.cn}}
\author{Zhao-Feng Ge\footnote{gezhaofeng@itp.ac.cn}}

\affiliation{CAS Key Laboratory of Theoretical Physics, Institute of Theoretical Physics,
Chinese Academy of Sciences, Beijing 100190, China\\
School of Physical Sciences, University of Chinese
Academy of Sciences, Beijing 100049, China}

\begin{abstract}
We investigate a dynamic model described by the classical Hamiltonian $H(x,p)=(x^2+a^2)(p^2+a^2)$, where $a^2>0$, in classical, semi-classical, and quantum mechanics.
  In the high-energy
  $E$ limit, the phase path resembles that of the $(XP)^2$ model. However, the non-zero value of $a$ acts as a regulator, removing the singularities that appear in the region where $x, p \sim 0$, resulting in a discrete spectrum characterized by a logarithmic increase in state density. Classical solutions are described by elliptic functions, with the period being determined by elliptic integrals. In semi-classical approximation, we speculate that the asymptotic Riemann-Siegel formula may be interpreted as summing over contributions from multiply phase paths. We present three different forms of quantized Hamiltonians, and  reformulate them into the standard Schr\" odinger equation with $\cosh 2x$-like potentials. Numerical evaluations of the spectra for these forms are carried out and reveal minor differences in energy levels. Among them, one interesting form possesses
  Hamiltonian in the Schr\" odinger equation that is identical to its classical version. In such scenarios, the eigenvalue equations can be expressed
  as the vanishing of the Mathieu functions' value at $i\infty$ points,
  and furthermore, the Mathieu functions can be represented as the wave functions.
\end{abstract}

\maketitle
\section{Introduction\label{sec1}}
The Hilbert-Polya conjecture posits that the non-trivial zeros of the Riemann zeta function align with the eigenvalues of a self-adjoint operator, thereby linking a mathematical conundrum to a physics challenge. While this conjecture presents a promising avenue to validate the Riemann Hypothesis (RH) %\cite{riemann:1},
identifying such a self-adjoint operator whose spectrum mirrors the imaginary part of the zeros of the Riemann zeta function, proves challenging.

Assuming  validation of RH,  Montgomery~\cite{montgomery} and Odlyzko~\cite{Odlyzko} derived that   asymptotic local statistical behaviour  of those Riemann zeros resembles Gaussian Unitary Ensemble (GUE) in random matrix model. Berry~\cite{berry:1986} suggested that there may exist certain classical Hamiltonian system with chaotic dynamics and isolated  periodic prime-number-related orbits and the spectrum of such corresponding quantum system may be reproduce the Riemann zeroes.
%In the several decades after it is proposed, the conjecture is supported by results and analogies %involving number theory, random matrix theory and quantum %chaos\cite{impr:1,impr:2,impr:3,impr:4,impr:5,impr:6,impr:7,impr:8,impr:9,impr:10,impr:11,impr:12,impr:1%3,impr:14,impr:15,impr:16,impr:17,impr:18}.\par

Significant progress was made by and Connes~\cite{connes:1998},  Berry and Keating~\cite{berry:1999} in 1999 when they proposed the $XP$ model.  The primary model is singular as
$x,p\sim 0$ and predicts a continuous spectrum.
Imposing constraints  on the $XP$ model in the phase spaces, Berry, Keating and Connes  derived  discrete spectra; notably, the state density scales with
$\log E$ at high energy limits in semi-classical estimations. This logarithmic term together with other finite terms  successfully reproduce the average density of the Riemann zero.

In quantum physics, this logarithmic association holds particular intrigue, potentially hinting at scaling violations. The allure of the regularized $XP$ models lies in their spectral behaviour, which resonates with that of the Riemann zeros.

While a full quantum description of the regularized Berry, Keating and Connes $XP$ model~\cite{berry:1999,connes:1998} is absent, modified $XP$ models with classical Hamiltonians, such as $H = x(p+1/p)$ proposed by Sierra and Rodr¨ªguez-Laguna~\cite{sierra:1}, and $H = (x+1/x)(p+1/p)$ proposed by Berry and Keating~\cite{berry:2011}, have been introduced.  The spectra predicted by these models are coincidence with the smooth approximation of the Riemann zeros, and later Sierra generalized the Hamiltonians in terms of the family of Hamiltonians $H = U(x)p+V(x)/p$~\cite{sierra:2012}, capturing the dynamics of a massive particle in a relativistic spacetime, with metrics informed by functions U and V. Subsequent works inspired by the $XP$ model have delved deeper into models associated with the Riemann zeros~\cite{bender:2017}. In all above models, the dynamic variable $p$ appears as nonlocal functions of $p$, which are accommodated in conventional quantum mechanics.

In this paper, we propose a $(XP)^2$-like model, with classical Hamiltonian   defined as
 $H=(x^2+a^2)(p^2+a^2)$.
 This model offers an alternative regularization to the traditional $XP$ model,  where parameter
 $a$ acts as a regulator removing the singularities for $x,p$ near zero, thus, a discrete spectrum is expected after imposing quantization.
Semi-classical estimation shows that  the state density is proportional to  $\log E$ in the high energy limit.
The
Hamiltonian   in our model is local functions of $x$ and $p$, thus
 quantization of this model is straightforward, although it suffers from  the operator ordering ambiguities. We explore multiple versions of the quantized Hamiltonian.
 Our numerical analyses depict minor spectral differences in high energy limit. Through variable transformations, we elucidate that both the classical Hamiltonian and its quantum counterparts can be framed within the standard Schr\"odinger equation with $\cosh^2$-like potentials. We furnish numerical comparisons across various versions and interpret their physical implications.
 As all single degree of models such as the regularized $XP$ model~\cite{connes:1998,berry:1999}, and those models presented in~\cite{sierra:1,berry:2011,sierra:2012}, our single degree of model can only
 reproduce the average density of the Riemann zeros, but not their local statistical behaviour, arising from the randomness in high energy limit.

Notice that all the statistical properties of the Riemann zeros can be fully understood by the Riemann-Siegel formula. To understand the physical implications of the Riemann-Siegel formula, which are essential for calculating the  Riemann zeroes, we speculate that there may be contributions from all possible phase paths by scaling $a\to na, n=1,2,\cdots$, with $n$ naturally truncated to the integer part of $\sqrt{E/2\pi}$  in the semiclassical approximation.

 This paper unfolds as follows: Section II elaborates on the classical solutions. Section III delves into the semi-classical implications on spectra.  Section IV explore  multiple versions of the quantized Hamiltonian and furnishes numerical calculations on spectra and wave functions. Section V bridges our findings with conventional Schr\" odinger equations. Lastly, Section VI contributes conclusions.

\section{
	Classical solutions\label{sec2}}
We start by introducing a classical Hamiltonian defined as:
\begin{equation}\label{H-c}
H=(x^2+a^2)(p^2+a^2)
\,
\end{equation}
where $a$ is real number. For $a=0$,  this Hamiltonian reduces to the square of the
 $XP$  model, and consequently, its dynamics closely resemble  those of the conventional
 $XP$ model. The non-zero
 $a$ acts as a regulator, constraining the system's path within a finite region in the phase space.

  For a more general form
\begin{equation}\label{H-c-g}
	H=(x^2+a^2)(p^2+b^2)
	\,
\end{equation}
with $a\neq b $, one may impose a canonical transformation to rescale $x$, $p$ as  $x'=r x$, $p'=p/r$ respectively, using a real number $r=\sqrt{|a|/|b|}$. This transformation allows us to rewrite the Hamiltonian in the form of equation Eq.(\ref{H-c}).
\begin{equation}\label{H-c-1}
	H=(x'^2+|ab|)(p'^2+|ab|)
	\,,
\end{equation}
which aligns with the form of  Eq.(\ref{H-c}).

\begin{figure}[!t]
\centering
\includegraphics[scale=0.6]{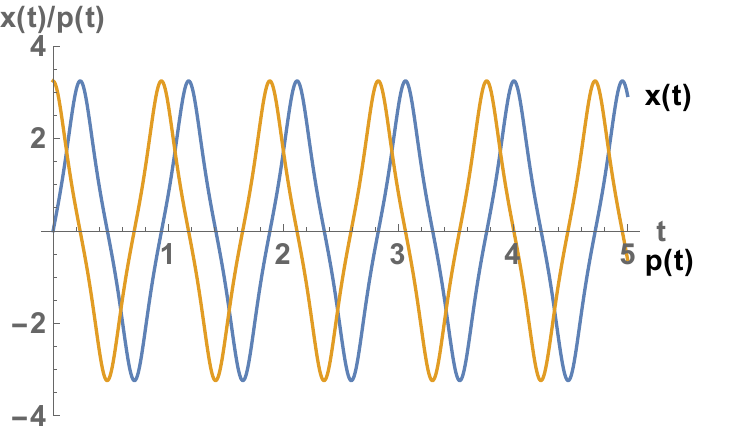}
\caption{Classical trajectory of $H = (x^2+a^2)(p^2+a^2)$, where $a=\sqrt{2}, E=5, x(t=0)=0$.}
\label{fig:1}
\end{figure}

\begin{figure}[!t]
\centering
\includegraphics[scale=0.6]{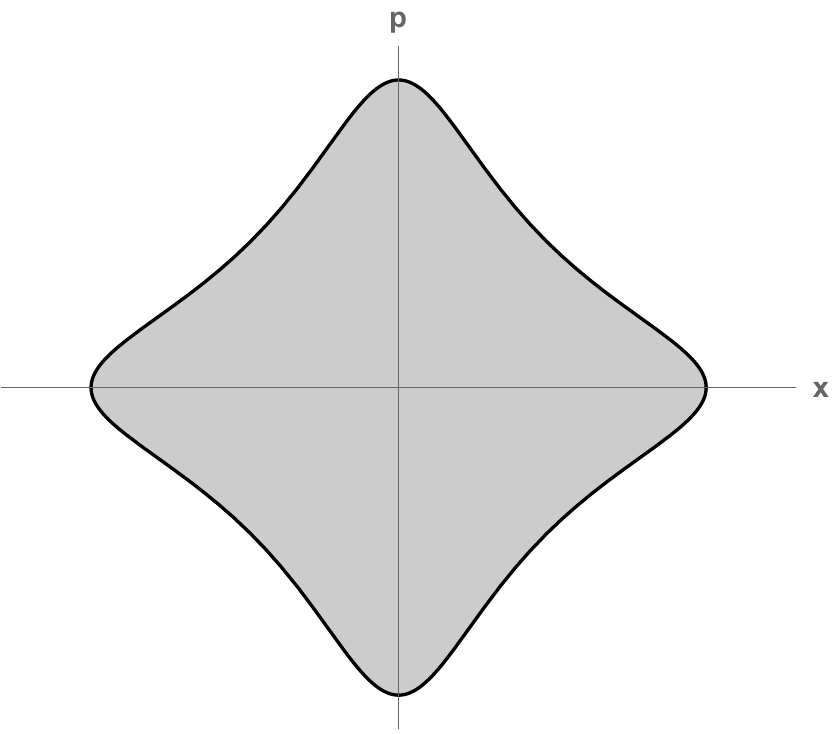}
\caption{The region between the two curves describes the allowed phase space with area A bounded by the classical path.}
\label{fig:2}
\end{figure}

The classical canonical Hamilton equations of motion are  given by
\begin{eqnarray}
\dot{x}&=&\frac{\partial H}{ \partial p}=2p(x^2+a^2)\;\;\;,\label{x-dot}\\
\dot{p}&=&-\frac{\partial H}{ \partial x} =-2x(p^2+a^2)\;\;\;.
\label{p-dot}
\end{eqnarray}
This system conserves energy, with paths in phase space adhering to a constant energy surface given by $H_E= (x^2+a^2)(p^2+a^2)$. For a specified energy $H_E$, $x$ and $p$ move periodically with respect to time
$t$, as depicted in~Fig.~\ref{fig:1}.   Setting $H_E=E^2+a^4$ , these equations simplify to:
\begin{eqnarray}
\dot{x}&=& \sqrt{(x^2+a^2)(E^2-a^2x^2)}\;\;\;,\\
\dot{p}&=&-\sqrt{(p^2+a^2)(E^2-a^2p^2)}\;\;\;,
\end{eqnarray}
in which  $x(t)$, $p(t)$ are decoupled and then  can be expressed by  the  elliptic functions~\cite{elliptic} ${\rm sn(u|m)}$
\footnote{the  elliptic function ${\rm sn(u|m)}$ is defined as the inverse function of the elliptic integral $u=\int^\phi_0 d\phi /\sqrt{1-m^2\sin^2\phi}$, via ${\rm sn(u|m)} \equiv \sin \phi$. It is double periodic function in the complex $u$ plane with $4  \mathcal {K}(m)$ and $2 \,i\mathcal {K}(1-m)$ periods, with $\mathcal {K}(x)$ being the first class complete elliptic integral defined in the context }
\begin{eqnarray}
x(t)&=&\frac{E}{a}\, {\rm sn}\left(2\,a^2\,t\Big|-\frac{E^2}{a^4} \right)\,, \\
p(t)&=&\frac{E}{a}\, {\rm sn}\left(2\,a^2\,\left( t +\frac{T_E}{4}\right)\Big| -\frac{E^2}{a^4} \right)\,,
\end{eqnarray}
where $T_E$ denotes the period time, characterized by the complete elliptic integral of the first kind,
   $\mathcal {K}(x)$:
\begin{equation}\label{Te}
T_E =  \frac{2}{a^2}{\displaystyle \int_{0}^{\frac{\pi}{2}}d\theta\,\frac{1}{\sqrt{1+\frac{E^2}{a^4}\sin^2\theta}}}
\equiv \frac{2}{a^2}\,\mathcal{K}\left(-\frac{E^2}{a^4}\right)\;.
\end{equation}
Here we have set initial condition $x(t=0)=0$.  The solution of $x(t)$ and $p(t)$ retain the same  form as expected but $\pi/4$ phase difference. In Fig.~\ref{fig:2}, the allowed phase space is shown, bounded by the classical path.

In the high energy limit, $E\to \infty$, the leading term of
$T_E$ goes to
\begin{equation}\label{Te-lim}
\lim_{E\to \infty} T_E = \lim_{E\to \infty}\frac{2}{a^2}\,{\cal K } \left(-\frac{E^2}{a^4}\right)=\frac{2}{E}\ln\left(\frac{4E}{a^2}\right)\;\,.
\end{equation}
Thus in this limit, $T_E$ exhibits a proportionality to the inverse power of
$E$ and the logarithmic factor
 $\ln\left(\frac{4E}{a^2}\right)$, which can
 be interpreted as scaling violation, and the parameter $a$
 acts as a regulator, ensuring
 $T_E$ remains finite. This result can be interpreted through Eq.~(\ref{x-dot}) . Actually, in the high energy limit,
\begin{equation}\label{x-dot-appx}
\dot{x} =2P(x^2+a^2)\simeq 2Px^2\simeq 2E\,x\;,
\end{equation}
from which we infer that :
\begin{equation}\label{T-E-sem}
T_E\simeq\frac{2}{E}\int_a^{E/a} \frac{dx}{x} \simeq\frac{2}{E} \ln\frac{E}{a^2}\;,
\end{equation}
which reproduce the logarithmic $E$ term in Eq.~(\ref{Te-lim}).
In the next section, we will see that this logarithmic factor relates to the state density logarithmic growth with the energy increasing it high energy limit.

Finally, we explore a classical canonical transformation from $x,p\to u,v $, which  preserves the classical Poisson bracket, represented as:
\begin{eqnarray}
x&=&a\,\sinh u \,,\label{x-u} \\
p&=&\frac{v}{a\,\cosh u} \label{p-uv}\,\,.
\end{eqnarray}
 Following the transformation, the Hamiltonian can be reframed as:
 \begin{equation}\label{H-uv}
 H(u,v)=v^2+a^4\cosh ^2u=v^2+a^4\sinh ^2u+a^4\;.
 \end{equation}

 This Hamiltonian corresponds to the standard form for an object with half-unit mass, moving one-dimensionally in the potential
  $a^4\cosh ^2u=a^4\sinh^2u+a^4 $.

  The findings indicate that, at the classical mechanics level, the Hamiltonians presented in Eq.~(\ref{H-c}) and  Eq.~(\ref{H-uv})  depict identical dynamics.

\section{	Semi-classical estimation for the spectrum\label{sec3}}

The semi-classical estimation serves as a bridge between classical and quantum descriptions, offering insights into fundamental features of quantum representation. We employ the semi-classical approximation to estimate the spectrum of the Hamiltonians presented in
(\ref{H-c}) and in (\ref{H-uv}) and to delve into their high-energy asymptotic behavior.

%trajectory of the Hamiltonian $H = (x^2+a^2)(p^2+a^2)$,

Considering the classical action variable:
\begin{equation}\label{I}
S(a)=\frac{1}{2\pi}\oint p dq,
\end{equation}
which is the counter area of the phase space trajectory with given energy $E$, over $2 \pi$.  This action variable remains invariant under classical canonical transformations and is a function of $E$.

The one-dimensional periodic motion suggests a discrete energy spectrum. Thus, the semi-classical approximation can be employed to estimate
 $N(E)$,
 the number of the quantum states below the energy $E$:
\begin{equation}
N(E) \equiv \frac{S(a)}{h}.
\end{equation}

Consequently, for the Hamiltonians defined in
  (\ref{H-c}) and in (\ref{H-uv}), we deduce:
\begin{equation}
N(E)= \frac{4}{h}\int_{0}^{\frac{E}{a}}\sqrt{\frac{E^2-a^2x^2}{x^2+a^2}} dx\;.
\end{equation}
This integral can be represented using the incomplete elliptic integral of the second kind
 $\mathcal {E} \left(  \phi |m\right)$:
\begin{equation}
N(E)=- \frac{4i}{h} \enspace E \mathcal {E} \left( i\sinh^{-1}\frac {E}{a^2} \Big| \frac{a^4}{E^2} \right)\;.
\end{equation}
It can also be defined using the complete elliptic integrals of the first kind
$\mathcal {K}(u)$ and the second kind $\mathcal {E}(u)$:
\begin{equation}
N(E)= \frac{4}{2\pi\hbar}\left[\frac{E^2+a^4}{a^2} \mathcal {K}\left(-E^2/a^4\right)-a^2 \mathcal{E}\left(-E^2/a^4\right) \right] \;,
\end{equation}
$\mathcal {E}(u)$ is defined as
\begin{equation}\label{CE-2}
	\mathcal {E}(u)=  {\displaystyle \int_{0}^{\frac{\pi}{2}}d\theta\,{\sqrt{1+\frac{E^2}{a^4}\sin^2\theta}}}
	\equiv \frac{2}{a^2}\,\mathcal{K}\left(-\frac{E^2}{a^4}\right)\;.
\end{equation}

The energy gap between two adjacent bound states can be approximated as:
\begin{equation}\label{E-s}
\Delta E = \hbar\frac{ d E }{ d I } \Delta n =  \frac{ \hbar }{ 2E } \frac{ d H }{ d I } = \frac{\omega_E \hbar }{ 2 E } = \frac{ \pi\hbar }{ T_E E }\;\,,
\end{equation}
where ${\omega_E}$ denotes the circular frequency for periodic motion at energy $E$. Referring to Eq.~(\ref{Te-lim}),    in high energy limit,
\begin{equation}\label{E-s-lim}
\Delta E = \frac{ \pi\hbar }{\displaystyle 2 \ln  \frac{ 4 E }{ a^2 }}
\;.
\end{equation}
Therefore, from this semi-classic estimation we see that
in the high-energy limit, the energy gap between two adjacent bound states is proportional to the inverse power of $\ln E$, as expected.

In the high-energy limit,

\begin{equation}\label{lim-N}
	\lim_{E\to\infty}N(E)= \frac{4}{2\pi\hbar}E\left(\log \left({4E\over a^2}\right)-1\right) \;.
\end{equation}

For models with $x,p$ exchanging symmetry, the state can be classified as 4
sectors according to the $x,p$ exchanging properites\footnote{For instance, for harmonic $H=(x^2+p^2)/2$, the wave functions in configuration space is proportional to that in momentum space by a factor $\lambda=1,i,-1,-i$, with $\lambda^4=1$. The states are classified into 4 sectors according to different values of $\lambda$}.
Considering Maslov index, precisely, each of these sectors satisfied the following state number counting rule:
\begin{equation}\label{lim-N-j}
	N'_j(E)={1\over4}\left(N(E)+j-{1\over 2 }\right)\;\;\;j=0,1,2,3
\end{equation}
Taking $j=3$ and $a^2=8\pi$,
\begin{equation}\label{lim-N-3}
	\lim_{E\to\infty}N'_3(E)=	 {E\over 2\pi }\left(\log \left({E\over 2\pi}\right)-1\right)+{7\over 8 }\;.
\end{equation}
 This number is exactly equal to that given by the constrained $XP$ mode~\cite{berry:1999}, and hence exactly the same with the asymptotic form of the smoothed counting function for the Riemann zeros.

We now try to interpret Riemann-Siegel formula in the semi-classical approximation by including contributions from multiple phase paths for given energy $E$. This is allowed in quantum physics according to path integral formula, where an object may stay multiple positions simultaneously, while in purely classical mechanics it is assumed that one object can only stay at a unique place at a given time. To this end,
we first rescale the constant $a  \to na$, with $n=1,2,\cdots$. For a fixed value of $E$, different values of $n$ correspond to different phase paths. As $n$ goes to larger and larger, the areas of $S(na)$ goes to zero. The maximal value of $n$ is then truncated to $[\sqrt{E\over 2 \pi}]$ ($[\cdots]$ means the integer part of the number). This value of $n$ is  the same with that truncated $n$ in  Riemann-Siegel formula  for the function $Z(E)$ defined by:
 \begin{equation}
 Z(E)=\exp\left(i\theta (E)\right)\zeta\left({1\over 2}+iE\right),
 \end{equation}
where
 \begin{equation}
\theta (E)=\arg \Gamma\left({1\over 4}+{1\over 2}iE\right)-\left({1\over 2} E \ln \pi\right),
\end{equation}
and
\begin{equation}
\zeta(s)=\sum_{n=1}^\infty\;{1\over n^s}\;,
\end{equation}
is the Riemann $\zeta$ function.

For a real $E$, real function $Z(E)$ can be expressed as Riemann-Siegel formula:
\begin{equation}
Z(E)=\sum_{n=1}^{\left[\sqrt{E\over 2 \pi}\right]} {\rm Re}\,\exp\left[i\left(\theta (E)-E\ln n\right)-{1\over 2}\ln n\right]+R(E),
\end{equation}
where $R(E)$ is the residue term which can be neglected in most cases.
In high energy $E$ limit,

\begin{equation}
Z(E)\simeq \sum_{n=1}^{\left[\sqrt{E\over 2 \pi}\right]} {\rm Re}\,
\exp\left[i\left({E\over 2}\ln {E\over 2\pi n^2}-{E\over 2}-{\pi\over 8}\right)-{1\over 2}\ln n\right]\;.
\end{equation}
This formula can be used to estimate almost all Riemann Zeroes expect for very few cases where Residue term can not be neglected.

Notice that Bohr-Sommerfeld quantization condition can also be expressed as
\begin{equation}\label{s0}
{\rm Re} \exp\left({i\,S(a)}\right)=0
\;.
\end{equation}

When we generalize this formula by  summing over  contributions from all  phase paths with allowed values of $n$
 \begin{equation}\label{sn}
 {\rm Re} \;\sum_{n=1}^{\left[\sqrt{E\over 2 \pi}\right]} \exp\left(i\,S(n\sqrt{8\pi})-{1\over 2} \ln n\right)=0
 \;.
  \end{equation}
 We found that it nicely reproduces the main part of the  Riemann-Siegel formula in the large $E$ limit. Here, the real term in the exponent may arise from quantum corrections. We emphasize here that this  empirical formula is only speculation, for we have not yet know how to derive it from a fundamental theory.

%%%%%%%%%%%%%%%%%%%%%%%%%%%%%%%%%%%%%%%%%%%

\section{Quantum descriptions \label{sec4}}
The quantum mechanics description for the models associated with the classical Hamiltonian presented in Eq.~(1) is introduced next. Through the process of canonical quantization, the classical variables
$x$ and
$p$ in the Hamiltonian are substituted with their operator counterparts
$\hat{x}$ and $\hat{p}$,
 which obey the  Poisson bracket $[\hat{x}, \hat{p}]=i\hbar$.
To simplify the expressions, we often set
 $\hbar=1$.
This resetting can be realized through the rescaling:
 $ \hat{x}\to\sqrt{\hbar}\hat{x} $, $ \hat{p}\to\sqrt{\hbar}\hat{p} $, $ a\to\hbar a $, $ E\to\hbar E $,  $ H\to\hbar^2 H $.
 The ensuing
 Poisson bracket  is  $[\hat{x}, \hat{p}]=i$.
  Notice that  classical Hamiltonian given in Eq.~(\ref{H-c}) contains multiplication of $x^2$ and $p^2$. Its quantization suffers from operator ordering ambiguities since $\hat{x} $ and $\hat{p}$ do not commutate with each other in quantum mechanics. This implies there exists different quantum Hamiltonian forms corresponding to the same classical Hamiltonian  Eq.~(\ref{H-c}). We enumerate several representative self-conjugate Hamiltonian forms below:

  Form I preserves the symmetry of
$ \hat{x},  \hat{p}$ exchange:
\begin{equation}\label{H-1}
	H_1(\hat{x},\hat{p})= \frac{1}{2}\left[\left( \hat{x}^2+a^2\right) \left( \hat{p}^2+a^2\right) +\left( \hat{p}^2+a^2\right) \left( \hat{x}^2+a^2\right)\right]  \;.
\end{equation}
This form can also be expressed as:
\begin{equation}\label{H-4}
	H'_1(\hat{x},\hat{p})=\frac{1}{2}\left( A{A}^\dagger+A^\dagger A\right) \, ,
\end{equation}
with
\begin{equation}\label{A}
	A \equiv \frac{1}{2}\left( \hat{x}\hat{p} +\hat{p}\hat{x} \right)+ia\left( \hat{x}+\hat{p}  \right) -a^2 \;,
\end{equation}
and
\begin{equation}\label{H-5}
	H''_1(\hat{x},\hat{p})=\frac{1}{2}\left( B{B}^\dagger+B^\dagger B\right) \,,
\end{equation}
with
\begin{equation}\label{B}
	B \equiv \frac{1}{2}\left( \hat{x}\hat{p} +\hat{p}\hat{x} \right)+ia\left( \hat{x}-\hat{p}  \right) -a^2 \;.
\end{equation}
The Hamiltonians $H'_1(\hat{x},\hat{p})$ and $H''_1(\hat{x},\hat{p})$ only differ from  $H_1(\hat{x},\hat{p})$ by a constant value:
\begin{equation}\label{H1-H4,5}
	H'_1(\hat{x},\hat{p})=H''_1(\hat{x},\hat{p})=H_1(\hat{x},\hat{p})
	+\frac{3}{4}\;.
\end{equation}
This implies that $H_1(\hat{x},\hat{p})$ $H'_1(\hat{x},\hat{p})$ $H''_1(\hat{x},\hat{p})$  describe identical quantum systems. Therefore, for Form I, the focus is primarily on  $H_1(\hat{x},\hat{p})$.

Form II, an alternative quantum version, is presented as:
\begin{equation}\label{H-2}
	H_2(\hat{x},\hat{p})= \sqrt[4]{\hat{x}^2+a^2}  \,\hat{p}\,\sqrt{\hat{x}^2+a^2}\,\hat{p}\, \sqrt[4]{\hat{x}^2+a^2}+a^2\left(\hat{x}^2+a^2\right)\;.
\end{equation}
Contrastingly, this form breaks $X,P$ exchange symmetry, yet it offers a more streamlined Schr\" odinger equation upon a variable transformation.

Yet another quantum Hamiltonian, denoted as Form III, is described as:
\begin{equation}\label{H-3}
	H_3(\hat{x},\hat{p})= \sqrt{\hat{x}^2+a^2} \left( \hat{p}^2+a^2\right) \sqrt{\hat{x}^2+a^2} \;.
\end{equation}
This form also breaks the $X,P$ exchange symmetry.

Given three kinds of quantum Hamiltonians above, one can derive differential equations for the wave-functions in the configuration space using the substitution  $\hat{x} \to  x$, $ \hat{p} \to  {\displaystyle -i\frac{d}{dx}} $.
\begin{equation}\label{Q-Wave}
H_I(x,-i\frac{d}{dx})\;\psi_I(x)=H_E\;\psi_I(x)
\;\;\;\;(I=1,2,3)\;.
\end{equation}
These differential equations serve as an eigenvalue equation  of self-conjugate linear operates. The solution of the $\psi_I(x)\;(I=1,2,3)$ is the eigenstate in Hilbert space with $H_E$ being the corresponding eigenvalue. The inner products of those wavefunctions are finite. This can be understood from the following observations.

Notice that in the $x\to \infty $ limit, keeping only the highest power terms of $x^2$, equations (\ref{Q-Wave}) are reduced to  $x^2\left(-{\displaystyle\frac{d^2}{dx^2}}+a^2\right)\psi_I(x)=0$, leading to the asymptotic behavior:
\begin{equation}\label{Wave-inf}
\psi_I(x) \;\stackrel{ x\to \infty }{\longrightarrow} \; e^{\pm |a x|}
\; \;\;\;\;(I=1,2,3) \;.
\end{equation}
For arbitrary real values of $H_E$, in the limit  $x\to \infty$,
$\psi_I(x)$ is a superposition of both
the convegent $e^{ -|a x|}$ term and the divergent $e^{ |a x|}$, hence is divergent.
When and only when $H_E$ is an eigenvalue, the
$e^{ -|a x|}$ term survives and  the $e^{ |a x|}$ term vanishes in the limit, i,e,
\begin{align}\label{wave-inf-pm}
\psi_I(x) \;\stackrel{ x\to \infty }{\longrightarrow} \; e^{-| a x|}\;\stackrel{ x\to \infty }{\longrightarrow}\;0 \;.
\end{align}
This asymptotic behavior of the eigenstates guarantees that  the inner products of those wavefunctions are finite and all the  eigenstates are defined in Hilbert space.

 %in Hilbert space%
It's evident that the Hamiltonians
 $ H_I(x)\;(I=1,2,3)$
remain invariant under parity transformations, i.e.
$H_I(\hat{x},\hat{p})=H_I(-\hat{x},-\hat{p})$. This suggests that physical wave functions possess definitive parities:

\begin{equation}\label{Wave-parity}
\psi_I(x) = \pm\; \psi_I(-x) \; \;\;\;\;(I=1,2,3) \;.
\end{equation}

With respect to all the aforementioned quantum Hamiltonians, Eq.~(\ref{Q-Wave}) can be restated as a secondary ordinary differential equation. Specific representations are:

For form I quantum Hamiltonian (\ref{H-1}),  the differential equation can be written as
\begin{equation}
(x^2+a^2)\;\psi_1''(x)+2x\;\psi_1'(x)+(E^2+1-a^2\,x^2)
\;\psi_1(x)=0 \;,\label{eq:1}
\end{equation}
here $H_E=E^2+a^4$ as in Sec. II.

For form II quantum Hamiltonian (\ref{H-2}),  the differential equation can be expressed as
\begin{equation}
\sqrt[4]{x^2+a^2}\;\frac{d}{dx}\left[\sqrt{x^2+a^2}\frac{d}{dx}\left(\sqrt[4]{x^2+a^2}\,\psi_{2}(x)\right)\right]+(E^2-a^2\,x^2)\;\psi_{2}(x)=0 \;.\label{eq:2}
\end{equation}
Setting $\phi_{2}(x)=\left(\sqrt[4]{x^2+a^2}\,\psi_{2}(x)\right)$, $\phi_{2}(x)$ then satisfies the following secondary ordinary differential equations
\begin{equation}
	(x^2+a^2)\;\phi_2''(x)+x\;\phi_2'(x)+(E^2-a^2\,x^2)\;\phi_2(x)=0 \label{eq:2-1}\;.
\end{equation}

For form III quantum Hamiltonian (\ref{H-3}),  the differential equation reads
\begin{equation}
	\sqrt{x^2+a^2}\;\frac{d^2}{dx^2}\left(\sqrt{x^2+a^2}\,\psi_{3}(x)\right)+(E^2+a^4)\;\psi_{3}(x)=0 \;.\label{eq:3}
\end{equation}
Setting $\phi_{3}(x)=\left(\sqrt{x^2+a^2}\,\psi_{3}(x)\right)$, it then satisfies the following secondary ordinary differential equations
\begin{equation}
	(x^2+a^2)\;\phi_{3}''(x)+(E^2-a^2\,x^2)\;\phi_{3}(x)=0 \;.
	\label{eq:3-1}
\end{equation}

Upon comparing Eqs.(\ref{eq:1}, \ref{eq:2-1}, \ref{eq:3-1}) corresponding to the distinct Hamiltonians (\ref{H-1}, \ref{H-2}, \ref{H-3}), it is evident that they can be consolidated into a general form:

\begin{equation}
	(x^2+a^2)\;\phi_I''(x)+(3-I)x\;\phi_I'(x)+(E^2+c_I
	-a^2\,x^2)\;\phi_I(x)=0\;\;\;\;\;\;(I=1,2,3)\;, \label{eq:123}
\end{equation}
assuming $\phi_1(x)=\psi_1(x)$  and $c_1=1,\;c_2= c_3=0$.

Employing the boundary conditions from Eq.~(\ref{wave-inf-pm}), one can numerically solve these equations to obtain eigenvalue
 $H_E=E^2+a^4$ and the associated wave functions.

The parity property in  Eq. (\ref{Wave-parity}) can streamline calculations.
For instance, parity-even solutions yield
\begin{equation*}
	\phi'(0)=0\;,
\end{equation*}%
while parity-odd solutions produce
\begin{equation*}
	\phi(0)=0\;.
\end{equation*}

\begin{figure}[!t]
	\centering
	\includegraphics[scale=0.4]{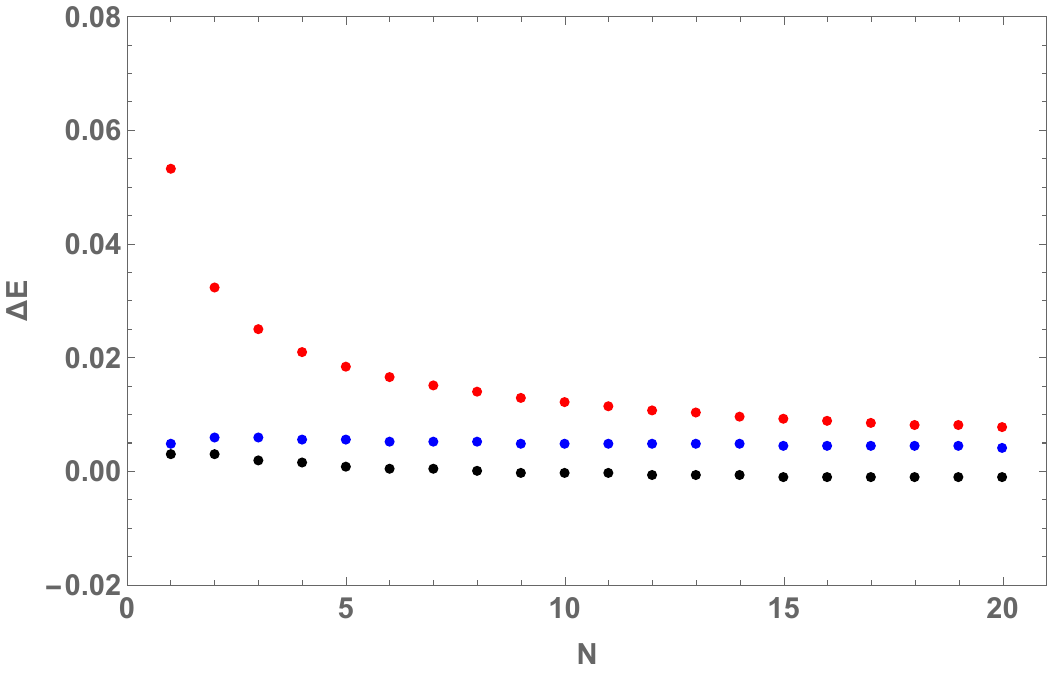}
	\caption{Energy differences $\Delta E$ between the eigenvalues predicted by $H_1$, $H_2$, and $H_3$ and that given by  semi-classical approximation, represented by  black, red, and blue points, respectively, with $a=\sqrt{8\pi}$.}
	\label{fig:6}
\end{figure}

Numerical analyses reveal that exact solutions closely align with semi-classical energy level estimations for $E$, especially as  $E\to \infty$.
We showcase the first 10 and the 40th-50th energy eigenvalues for $a=\sqrt{8\pi}$  in Table~\ref{table:1}, and in Table~\ref{table:2}, respectively. The energy differences $\Delta E$   between the eigenvalues predicted by $H_1$, $H_2$, and $H_3$ and that given by  semi-classical approximation for the first 20 energy levels, are shown in Fig.~\ref{fig:6}.

%
%The first 53 energy eigenvalues when a=1 are presented in \ref{fig:3}.
%\begin{figure}[!t]
%	\centering
%	\includegraphics[scale=0.6]{quan1.pdf}
%	\caption{Energy eigenvalues of $H=\frac{1}{2} [(x^2+a^2)(p^2+a^2)+(p^2+a^2)(x^2+a^2)]$, when $a=1$.}
%	\label{fig:3}
%\end{figure}
%\begin{figure}[!t]
%	\centering
%	\includegraphics[scale=0.6]{quan2.pdf}
%	\caption{Energy eigenvalues of $H=\sqrt{x^2+a^2}(x^2+a^2)\sqrt{x^2+a^2}$, when a=1.}
%	\label{fig:5}
%\end{figure}

%\begin{figure}[!t]
%	\centering
%	\includegraphics[scale=0.4]{n=1.pdf}
%	\includegraphics[scale=0.3]{n=2.pdf}
%	\includegraphics[scale=0.3]{n=50.pdf}
%	\includegraphics[scale=0.3]{n=51.pdf}
%	\caption{The Eigenfunctions of $H=\sqrt{x^2+a^2}(x^2+a^2)\sqrt{x^2+a^2}$ when n=1, 2, 50 and 51.}
%	\label{fig:6}
%\end{figure}

\begin{table}[htb]
\begin{center}
\caption{The first ten energy eigenvalues of $E$ predicted by $H_1$, $H_2$, $H_3$, and the semi-classical estimation, where $a=\sqrt{8\pi}$.}
\label{table:1}
\resizebox{\linewidth}{!}{
\begin{tabular}{|c|c|c|c|c|c|c|c|c|c|c|}

\hline    \diagbox{Hamiltonian}{N} & 1& 2&3&4&5& 6& 7 & 8&  9 &  10\\
\hline   H1 & 4.98755&8.72429&11.32743&13.47008&15.34559&17.04177& 18.60721& 20.07206 &  21.45654& 22.77496 \\
\hline   H2 & 5.03789& 8.75368& 11.35044& 13.48970&15.36304& 17.05767& 18.62192& 20.08583& 21.46954&22.78732 \\
\hline   H3 & 4.98943 & 8.72741 & 11.33123 &13.47432 &15.35015 &17.04656& 18.61216& 20.07713 &21.46169 &22.78018\\
\hline   Semi-classical & 4.98463 & 8.72139 &11.32535  &13.46862 &15.34460 & 17.04115& 18.60687 &20.07196& 21.45664& 22.77524\\
\hline
\end{tabular}
}
\end{center}
\end{table}

\begin{table}[htb]
\begin{center}
\caption{The 40th-50th ten energy eigenvalues of $E$ predicted by $H_1$, $H_2$, $H_3$, and the semi-classical estimation, where $a=\sqrt{8\pi}$.}
\label{table:2}
\resizebox{\linewidth}{!}{
\begin{tabular}{|c|c|c|c|c|c|c|c|c|c|c|}
%\begin{tabular}{|m{1.2cm}<{\centering}|m{1.2cm}<{\centering}|m{1.2cm}<{\centering}|m{1.2cm}<{\centering}|m{1.2cm}<{\centering}|m{1.2cm}<{\centering}|m{1.2cm}<{\centering}|m{1.2cm}<{\centering}|m{1.2cm}<{\centering}|m{1.2cm}<{\centering}|m{1.2cm}<{\centering}|}
\hline    \diagbox{Hamiltonian}{N} & 41& 42&43&44&45& 46& 47 & 48&  49 &  50\\
\hline   H1 & 51.73256&52.49623&53.25426&54.00686&54.75421&55.49649&56.23388&56.96652&57.69458&58.41818 \\
\hline   H2 & 51.73880&52.50239&53.26035&54.01289&54.76017&55.50240&56.23972&56.97231&57.70030&58.42385 \\
\hline   H3 & 51.73673&52.50035&53.25833&54.01089&54.75819&55.50043&56.23777&56.97037&57.69838&58.42194\\
\hline   Semi-classical& 51.73402 & 52.49769 &53.25572 &54.00832 &54.75568& 55.49797& 56.23535& 56.96800& 57.69605&58.41965\\
\hline
\end{tabular}
}
\end{center}
\end{table}

\section{Schr\" odinger forms \label{sec5}}

In this section, we demonstrate that all quantum  Hamiltonians  specified in Eqs. (\ref{H-1}), (\ref{H-2}), (\ref{H-3}) can be  recast  into a standard Schr\" odinger form. In this form, the Hamiltonian is  expressed as a sum
of kinetic term and a potential term.
While the kinetic terms consistently remain the same, the potential components exhibit variations.

We first impose a variable substitution, Eq.~(\ref{x-u}), $x=a\sinh u$ on the  the differential equations (\ref{eq:1})  (\ref{eq:2-1})  (\ref{eq:3-1}). The derivative term then  expressed as:
\begin{equation}
	\frac{d\phi}{dx} = \frac{d u}{dx}\;\frac{d\phi}{du}=\frac{1 }{a\cosh u}\;\frac{d\phi}{du}\;. \label{D-u}
\end{equation}
With (\ref{x-u}) and (\ref{D-u}), Eq.(\ref{eq:1}) can be rewritten as:
 \begin{equation}
	\psi_1''(u)+\tanh u\,\psi_1'(u)
	+(E^2+1-a^4\sinh^2 u)\psi_1(u)=0\;.\label{eq:1-2}
\end{equation}
Giving   $\varphi_1(u)= \sqrt{\cosh(u)}\;\psi_1(u)$, it reduces to   the following differential equation:
\begin{equation}
	\varphi_1''(u)+\left(\frac{1}{2}+E^2+\frac{1}{4}\tanh^2 u-
	a^4\sinh^2 u\right)
	\varphi_1(u)=0\; ,\label{eq:1-3}
\end{equation}
where the second derivative term is preserved but the first derivative term is eliminated. Consequently,
 $\varphi_1(u)$ adheres to the standard Schr\" odinger equation with a half unit mass and a characteristic effective potentia
\begin{equation}
	V_1(u)=-\frac{1}{2}-\frac{1}{4}\tanh^2 u+a^4\sinh^2 u\;.\label{V-1}
\end{equation}
We then derive the   Schr\" odinger  form   of the Hamiltonian form (\ref{H-2}).  By applying Eqs.(\ref{x-u},\ref{D-u}), Eq.(\ref{eq:2-1}) can be rewritten as:
\begin{equation}
	\psi_2''(u)
	+(E^2-a^4\sinh^2 u)\psi_2(u)=0\;.\label{2-2}
\end{equation}
Therefore, $\psi_2(u)$ complies with the standard Schr\" odinger equation, once again sporting a half unit mass but associated with an alternative effective potential:
\begin{equation}
	V_2(u)=a^4\sinh^2(u)\;.\label{V-2}
\end{equation}
We emphasis here  that this Schr\" odinger Hamiltonian remains exactly the same form as the classic one Eq.~(\ref{H-uv}).

Lastly, addressing the Hamiltonian form from Eq.~(\ref{H-3}) and applying Eqs.~(\ref{x-u},\ref{D-u}),  Eq.~(\ref{eq:3-1}) can be cast into:
\begin{equation}
	\phi_3''(u)-\tanh u\,\phi_3'(u)
	+(E^2-a^4\sinh^2 u)\phi_3(u)=0\;.\label{3-2}
\end{equation}
Introducing   $\varphi_3(u)= \phi_3(u)/\sqrt{\cosh u}$, it conforms to the differential equation:
\begin{equation}
	\varphi_3''(u)+\left(\frac{1}{2}+E^2-\frac{3}{4}\tanh^2u-
	a^4\sinh^2u\right)
	\varphi_3(u)=0\; .\label{3-3}
\end{equation}
Again, the second derivative term is retained while the first one vanishes. This ensures that
$\varphi_3(u)$ aligns with the standard Schr\" odinger equation, furnished with half unit mass, and steered by its own unique effective potential:
\begin{equation}
	V_3(u)=-\frac{1}{2}+\frac{3}{4}\tanh^2u+a^4\sinh^2 u\;.\label{V-3}
\end{equation}

\begin{figure}[!t]
\centering
\includegraphics[scale=0.6]{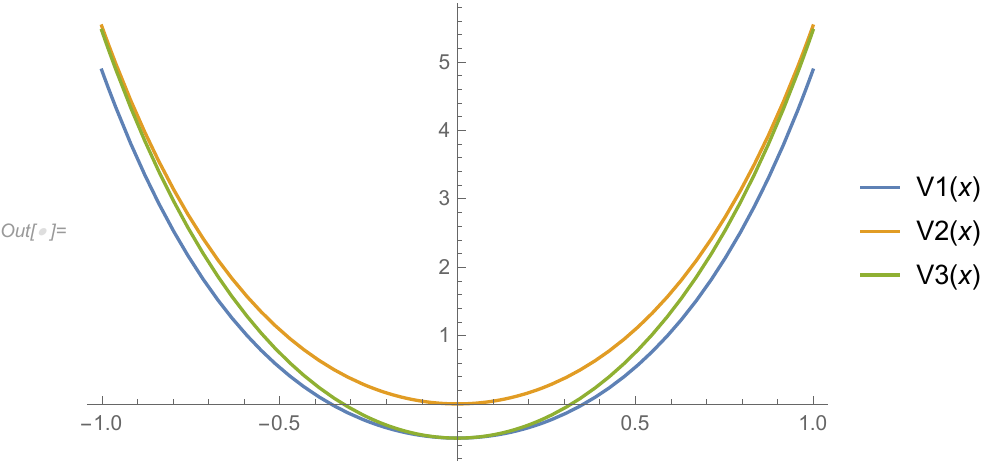}
\caption{The effective potentials $V_1(u)$,$V_2(u)$, and $V_3(u)$ with $a=\sqrt{2}$.}
\label{fig:4}
\end{figure}

We show three kinds of effective potentials $V_1(u),V_2(u)$ and $V_3(u)$ in Fig.\ref{fig:4} ,when $a = \sqrt{2}$.
Obversely,  those   Schr\" odinger forms differ  only by the potentials.   For
all value of, the differences among them are minor as shown in Fig.\ref{fig:4}.   Especially,
when $u\to \pm\infty$, all potentials $V_1(u)$, $V_2(u)$, and $V_3(u)$ given in Eqs.~(\ref{V-1})(\ref{V-2})(\ref{V-3}), respectively,  exponentially increase   with $u$:   $e^{2 |u|}$, with  constant differences.

Such Schr\" odinger equations can be tackled as eigenvalue problems. For an arbitrary real number
 $E$,  these solutions are divergent when $u\to \pm \infty $.
 Actually,  as discussed in the last section in Eq.~(\ref{Wave-inf}), when $u\to \pm \infty $, following either  solutions are possible:
\begin{equation}\label{Wave-u-inf}
\varphi_I(u) \;\stackrel{ u\to \infty }{\longrightarrow} \; e^{\pm a^2\cosh u}\;\;\;\;(I=1,2,3) \;.
\end{equation}
 When and only when $H_E=E^2+a^4$ is the eigenvalue, the following convengent solution  survives
\begin{align}\label{wave-u-inf-pm}
\varphi_I(u) \;\stackrel{ u\to \pm \infty }{\longrightarrow} \;e^{- a^2 \cosh u}\;{\longrightarrow} \;0\;\; \;(I=1,2,3) \;,
\end{align}
Then the solution as the wave-function is normalizable in Hilbert space.

Given the parity invariance of these Hamiltonians, eigenstates are distinctly marked by definitive parities, either parity-odd or parity-even. Implementing this condition alongside boundary condition:
\begin{align}\label{wave-u-bound}
	\varphi_I(\infty)=0 \;\; \;(I=1,2,3) \;,
\end{align}
one can numerically resolve the Schr\" odinger equations, procuring both eigenvalues and wave-functions.

We emphasize here that for Eq.~(\ref{2-2}),  solutions can be  gained  by  Mathieu functions. Original Mathieu functions serve as solutions for the differential equation:
\begin{equation}\label{Mathieu}
\frac{d^2\,y}{d\,w^2} +(c-2q \cos(2w))y=0\;,
\end{equation}
with parameters $c,q$, and variable $w$ all being real numbers. It is nothing but  the Schr\" odinger equation with unit mass and periodic potential $2\cos 2w$. Two independent solutions exist:
 $y_1(w)={\cal M }c(c,q,w)$
and $y_2(w)={\cal M }s(c,q,w)$, corresponding to parity-even and parity-odd solutions respectively.

Notice that Eq.~(\ref{3-3}) are related to  Eq.~(\ref{Mathieu}),  by an analytic continuation via  setting variable $w=i u$ and parameters  $a^4=q$, $E^2=c$, resulting in  their solutions are connected by the same analytic continuation.

Specifically,   the parity odd and parity even solutions for Eq. (\ref{2-2}) are given by
 \begin{gather}
	\label{psi-2s}
	\varphi_2^-(u)={\mathcal M}s(-E^2-\frac{a^4}{2},-\frac{a^4}{4}, iu)\;,\\
	\varphi_2^+(u)={\mathcal{M}c}(-E^2-\frac{a^4}{2},-\frac{a^4}{4}, iu)\label{psi-2c}\;,
\end{gather}
 respectively.

\begin{figure}[!t]
\centering
\includegraphics[scale=0.4]{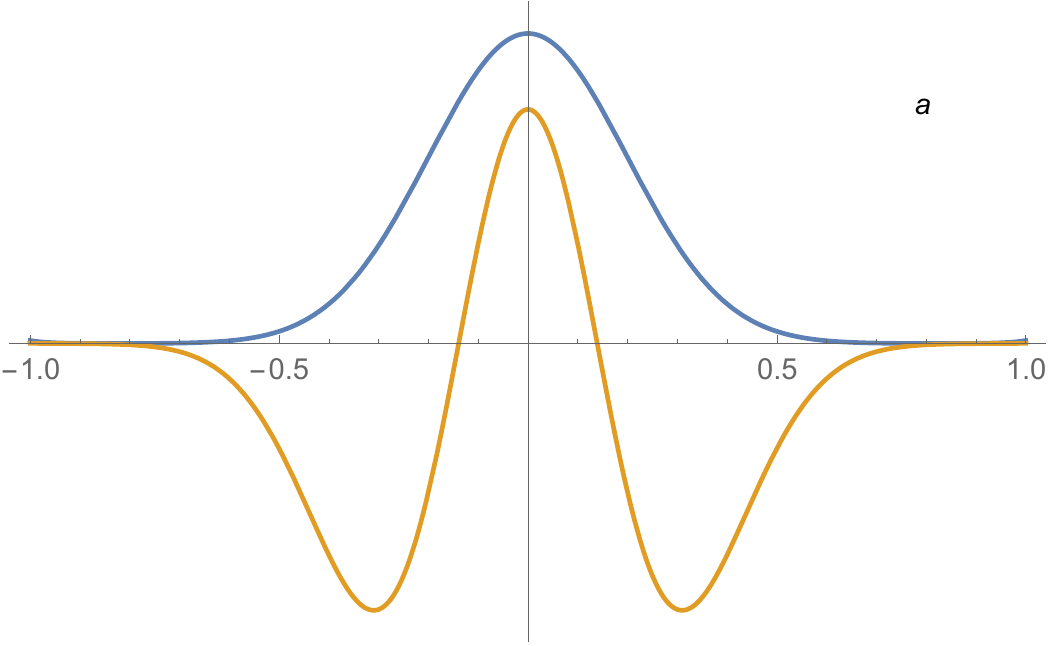}
\includegraphics[scale=0.4]{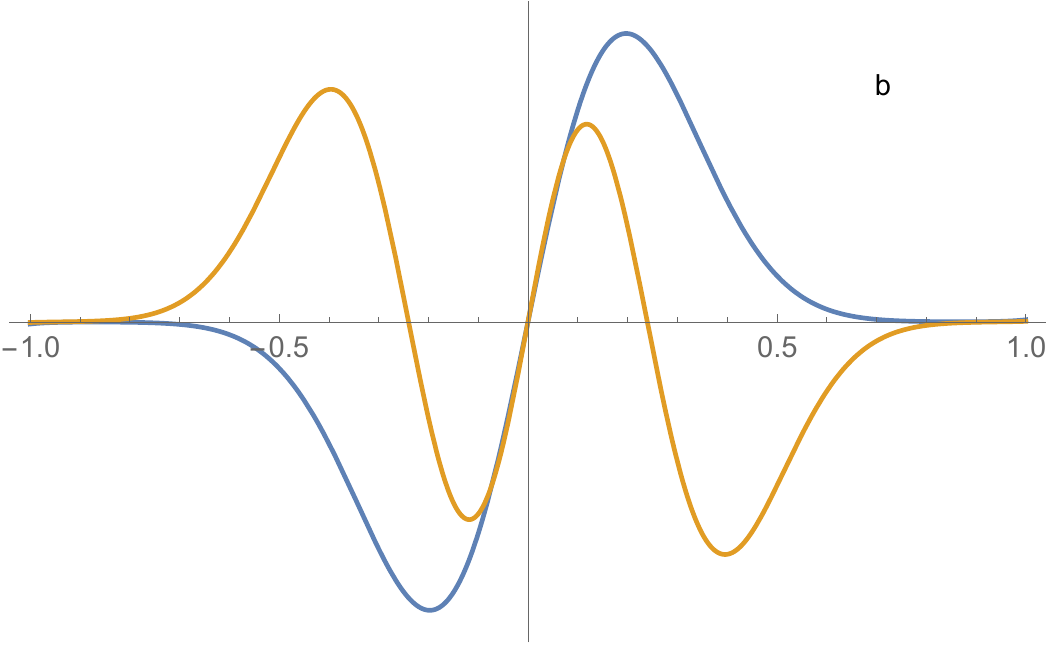}
\caption{The wavefunctions given by the Mathieu functions. (a) the first two parity odd eigenfunctions and (b) the first two parity even eigenfunctions, where $a=\sqrt{8 \pi}$.}
\label{fig:5}
\end{figure}

We then can apply boundary conditions Eq.~(\ref{wave-u-bound}),  expressed in the following forms
 \begin{gather}
 \varphi_2^- (\infty )= {\mathcal M}s(-E^2-\frac{a^4}{2},-\frac{a^4}{4}, i\infty)=
 \varphi_2^+(\infty)={\mathcal{M}c}(-E^2-\frac{a^4}{2},-\frac{a^4}{4}, i\infty)=0\;,
 \label{m-sc-bound}
 \end{gather}
to determine the  eigenvalues $H_E=E^2+a^4$.  Therefore they represent  the eigenvalue equations.  The solutions $\varphi_2^-(u )={\mathcal M}s(-E^2-\frac{a^4}{2},-\frac{a^4}{4}, iu) $ and $\varphi_2^+ (u )={\mathcal M}s(-E^2-\frac{a^4}{2},-\frac{a^4}{4}, iu)$ are the corresponding wavefunctions for parity odd and parity even, respectively.
We show the first two parity even eigenfunctions in Fig.~\ref{fig:5}(a) and the first two parity odd eigenfunctions in Fig.~\ref{fig:5}(b), where $a=\sqrt{8 \pi}$ and the corresponding eigenvalues can be read in Table~\ref{table:1}.

\section{conclusions}

In this paper, we have studied a dynamic model with the classical Hamiltonian   $H(x,p)=(x^2+a^2)(p^2+a^2)$ where $ ( a^2>0 )$   within the frameworks of  classic, semi-classic, and quantum mechanics, respectively. At high energy  $E$ limit, the path in phase space resembles that of the $(XP)^2$  model.

Historically, the
$XP$ model has been heralded as an intriguing paradigm that may potentially shed light on certain facets of the Riemann Hypothesis, especially when contextualized within the purview of the Hilbert-Polya conjecture. The primary $XP$  model yields a continuous spectrum. However, with appropriate regulation, the model exhibits a discrete spectrum. This spectrum is marked by a state density that increases logarithmically, analogous to the distribution of the Riemann zeros in the semi-classical approximation. We also speculate that asymptotic  Riemann-Siegel formula may be interpreted as summing over contributions from multiply phase paths.
A comprehensive quantum depiction of such a regulated
$XP$ model is conspicuously  absent in the existing literature. In this paper, we introduce a regulator for the
$(XP)^2$
model. Our work sheds some lights on  the quantization of the regulated $XP$  model, and further work on this subject is currently in preparation.

 We have explored three versions of the quantization  model and conducted numerical calculations for each.  In presenting
 their corresponding Schr\" odinger forms,  we found that quantum
 Hamiltonian form II given in (\ref{H-2}) is particularly intriguing for its
 Hamiltonian  appearing in Schr\" odinger Eq.~(\ref{eq:1-2}) is  exactly the  same  with the classic version Eq.~(\ref{H-uv}). In such cases, the eigenvalue equations can be expressed as conditions for the vanishing of the Mathieu functions' values at $i\infty$ points.
 Furthermore, these Mathieu functions can be presented as the wavefunctions of the system.

%\end{CJK*}

\end{document}